\documentclass[conference]{IEEEtran}
\IEEEoverridecommandlockouts
\usepackage{cite}
\usepackage{amsmath,amssymb,amsfonts}
\usepackage{algorithmic}
\usepackage{graphicx}
\usepackage{textcomp}
\usepackage{xcolor}
\usepackage{soul}
\usepackage{diagbox}
\usepackage{colortbl}
\usepackage{hhline}
\usepackage{float}
\usepackage{slashbox}
\usepackage[colorinlistoftodos,prependcaption]{todonotes}
\usepackage{hyperref}
\usepackage{subcaption}
\usepackage{pgfplots}
\usepackage{eurosym}
\usepackage{url}
\usepackage{color}
\usepackage{footmisc}
\usepackage{multirow}
\usepackage[utf8]{inputenc}
\usepackage{amsmath}
\usepackage{tikz}
\usetikzlibrary{patterns}

\usepackage{multicol}
\usetikzlibrary{shapes,arrows}
\usetikzlibrary{intersections}
\usepgfplotslibrary{units}
\usepackage{verbatim}
\usepackage{calc}
\usepackage{ifthen}
\usepackage{siunitx}
\usepackage[labelsep=period]{caption} %for dot after graphs
\usepackage{enumitem}
\usepackage{lipsum}
\usetikzlibrary{positioning}
\usetikzlibrary{calc}
\usepackage[export]{adjustbox}
\usepackage{delarray}
\floatstyle{plaintop}
\restylefloat{table}
\definecolor{bblue}{HTML}{4285F4}%4F81BD
\definecolor{rred}{HTML}{DB4437}%C0504D
\definecolor{ggreen}{HTML}{0F9D58}%9BBB59
\definecolor{yyellow}{HTML}{F4B400}%9F4C7C

\begin{document}

\title{Cloud Energy Micro-Moment Data Classification: A Platform Study
}

\makeatletter
\newcommand{\linebreakand}{%
  \end{@IEEEauthorhalign}
  \hfill\mbox{}\par
  \mbox{}\hfill\begin{@IEEEauthorhalign}
}
\makeatother
\author{
\IEEEauthorblockN{Abdullah Alsalemi, Ayman Al-Kababji,\\Yassine Himeur, Faycal Bensaali\\}
\IEEEauthorblockA{\textit{Department of Electrical Engineering} \\
\textit{Qatar University}\\
Doha, Qatar \\
\{a.alsalemi, aa1405810, yassine.himeur, f.bensaali\}@qu.edu.qa}
\and
\IEEEauthorblockN{Abbes Amira}
\IEEEauthorblockA{\textit{Institute of Artificial Intelligence} \\
\textit{De Montfort University}\\
Leicester, UK \\
abbes.amira@dmu.ac.uk}
}

\maketitle

\begin{abstract}
Energy efficiency is a crucial factor in the well-being of our planet. In parallel, Machine Learning (ML) plays an instrumental role in automating our lives and creating convenient workflows for enhancing behavior. So, analyzing energy behavior can help understand weak points and lay the path towards better interventions. Moving towards higher performance, cloud platforms can assist researchers in conducting classification trials that need high computational power. Under the larger umbrella of the Consumer Engagement Towards Energy Saving Behavior by means of Exploiting Micro Moments and Mobile Recommendation Systems (EM)\textsuperscript{3} framework, we aim to influence consumers' behavioral change via improving their power consumption consciousness. 
In this paper, common cloud artificial intelligence platforms are benchmarked and compared for micro-moment classification. Amazon Web Services, Google Cloud Platform, Google Colab, and Microsoft Azure Machine Learning are employed on simulated and real energy consumption datasets. The KNN, DNN, and SVM classifiers have been employed. Superb performance has been observed in the selected cloud platforms, showing relatively close performance. Yet, the nature of some algorithms limits the training performance.
\end{abstract}

\begin{IEEEkeywords}
cloud, data classification, energy efficiency, real-time, dataset, platform, study
\end{IEEEkeywords}

\section{Introduction}
\label{sec:introduction}
The European Union (EU) energy policy has considered energy efficiency as one of it its main targets \cite{dupont_defusing_2020}. By the Directive 2012/27/EU of 25 October 2012, the over arching goal is to accomplish the 2020 targets by the member states \cite{langsdorf2011eu}. Directive 2012/27/EU was revised to boost the energy efficiency of existing buildings, the ones in construction phase, and to re-emphasize on the energy performance of new upcoming buildings \cite{himeur2020efficient}.
From the technology aspect, the weight of factors accounting for the global market has been varying. Currently in 2020, control systems represent the largest technological portion of 21\% contributing in the global market. On the other hand, communication networks contribute with a share of 18\% after representing the largest portion of 20\% in 2012 \cite{Himeur2020IJIS-NILM}. Other technology aspects including field equipment, sensors, software, and hardware currently account for 44\% of the market; slightly dropping from 46\% in 2012 \cite{HIMEUR2020115872}.
Several areas in Information and Communications Technology (ICT) were investigated by Heras and Zarli \cite{de2008smart} to unlock potentials for the improvement of energy efficiency \cite{Varlamis2020CCIS}. These ICT areas include interoperability, building automation, and tools design and simulation. However, the areas of smart metering, user awareness, and decision support have been largely considered in recent research \cite{moranreducing, hannus2010ict, de2008smart,ye2008ict, horner2016known}, which emphasize on the significance of ICT in these areas. Smart metering is evident to be promising and technically practical throughout a variety of projects concluded across Europe, USA, and some other countries \cite{himeur2020improving}. By the means of information, rewards, and automation, Information Technology (IT) services can be integrated with metering infrastructure to enhance energy efficiency. Nevertheless, it is consumption awareness that is even held as more interesting for technology development. Smart metering data, including monitors through the internet on web applications and/or mobile devices, are made available to provide energy information and feedback tools \cite{hannus2010ict,Sardianos2020IJIS-ERS}.

From the data analysis aspect, more suitable behavioral interventions can be achieved throughout carefully examining the profile of a given consumer, with full details, to infer better conclusions \cite{Sardianos2020GreenCom}. Therefore, this work proposes the micro-moment concept as a novel scheme to analyze the daily segments of energy consumption with time-based and contextual snapshot \cite{alsalemi2019ieeesystems, ALSALEMI2019classifier}. Given a specific point in time, the power consumption of an appliance, annexed with other added information such as user preferences, constitute an energy micro-moment. Fig. \ref{fig:mm} illustrates an example of an energy micro-moment.

\begin{figure}[!ht]
\centering
\includegraphics[trim={0.8in 0.8in 0.9in 1in},clip,width=1\linewidth]{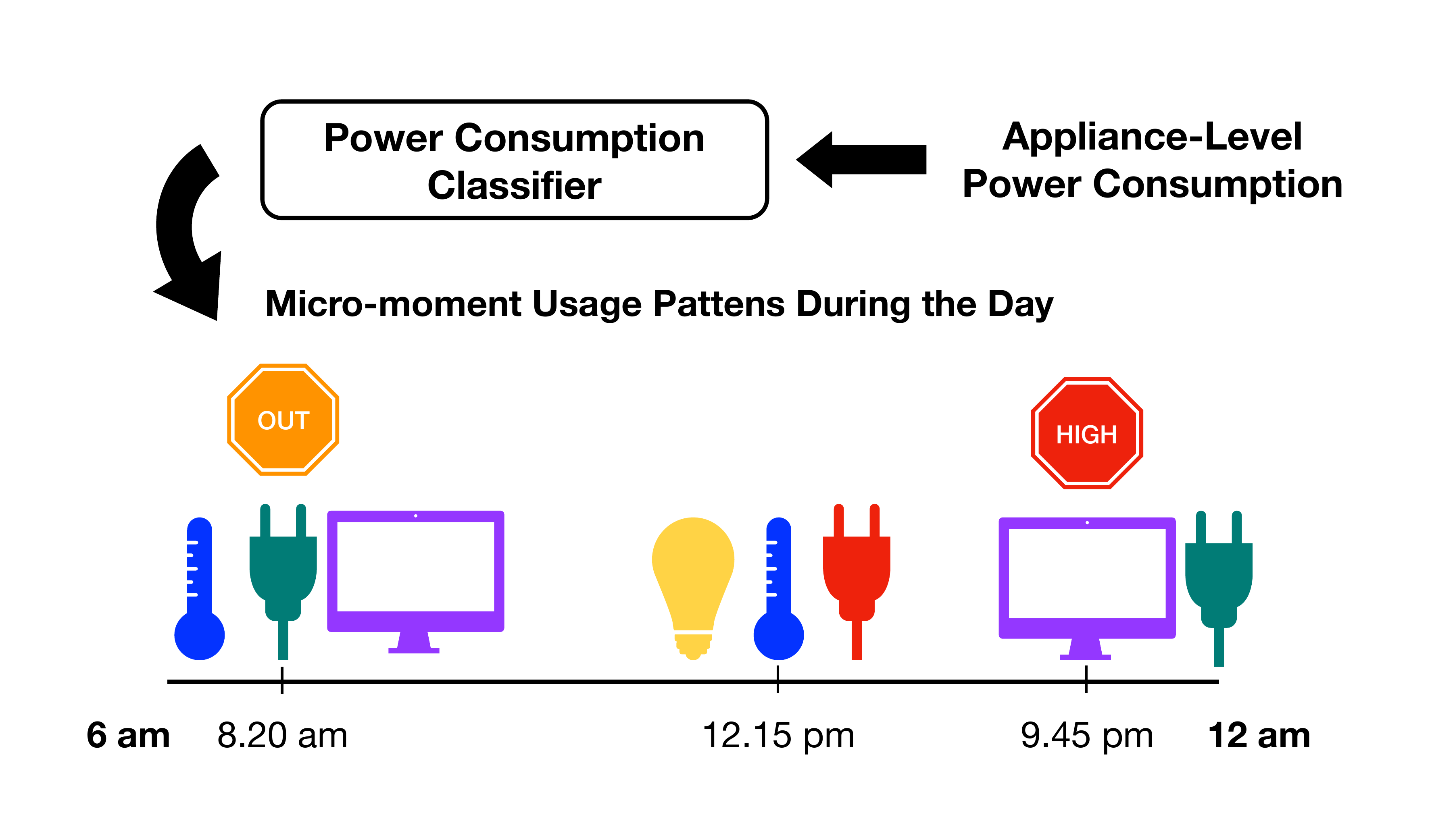}
\caption{An overview of a micro-moment example.}
\label{fig:mm} 
\end{figure}

In the field of health monitoring applications, Patel et al. \cite{patel2016wearable}. have addressed developing cloud-based ML models through a wearable computing platform. The ML pipeline is deployed to continuously evaluate the model’s performance, such that a degradation in performance can be detected. The model’s performance is evaluated with a recall and F1 score higher than 96\%, an overall recognition accuracy of 99.44\%, and a resting state model accuracy of 99.24\%. However, the accuracy is subject to limitations based on constrained settings of the collected data.

Bihis and Roychowdhury have adopted Microsoft Azure ML Studio as the cloud-based computing platform to implement a new generalized flow \cite{bihis2015generalized}. Through this generalized flow, the overall classification accuracy is maximized due to its ability of fulfilling multi-class and binary classification functions. The work also proposes a customized generalized flow of unique modular representations. The proposed approach is tested on three public datasets in contrast with existing cutting-edge methods, and results showed a classification accuracy of 78-97.5\%.

Chourasiya et al. have also adopted Microsoft Azure Machine Learning cloud, but for the classification of cyber-attacks \cite{roychowdhury2016ag}. The framework adopts a simple ML model with slight alteration, and by adjusting the multicast decision forest model, the results show an accuracy of 96.33\%. 

In this paper, we focus on the data processing aspect of micro-moments, particularly when cloud platforms are utilized as the computation engine. In the literature and commercial market, there is a wide pool of cloud ML services. While their features vary, many of cloud solutions include a free plan to allow researchers to get a taste of the power of cloud-based ML prior committing any financial investments.

The remainder of this paper is organized as follows. Section \ref{sec:data-collection-system} reviews the larger energy efficiency framework on which this work is based. Section \ref{sec:platforms} discusses evaluated cloud platforms. Sections \ref{sec:datasets} and \ref{sec:algorthims} reviews used datasets and the classification algorithms, respectively. Results are reported and discussed in Section \ref{sec:results}. The paper is concluded in Section \ref{sec:conclusion}.

\section{Overview of the (EM)\textsuperscript{3} Framework} \label{sec:data-collection-system}

The (EM)\textsuperscript{3} platform has been designed for two target user groups \cite{alsalemi_access_2020}: 
\begin{enumerate}
\item Homeowners that wish to reduce their energy footprint by avoiding unnecessary energy consumption, and by taking advantage of better energy tariffs that promote off-peak hours appliance usage; and
\item Office buildings that focus on the deactivation of unused appliances (e.g. monitors, lights, heating, and cooling devices, etc.) when weather conditions and room occupancy permits.
\end{enumerate}

The (EM)\textsuperscript{3} framework has been designed to support consumers behavioral change via improving power consumption consciousness. It includes four main steps defined as: collecting data (i.e. consumption footprints and ambient conditions) from different appliances in domestic buildings \cite{alsalemi_rtdpcc_2019, alsalemi2020micro}, processing consumption footprints in order to abstract energy micro-moments to detect abnormalities, deploying users' preferences information to detect the similarity amongst them \cite{himeur2020novel, himeur2020applicability, himeur2020data, himeur2020robust}, and generating personalized recommendations to reduce energy wastage based on a rule-based recommender model \cite{sardianos_smartgreens_2019, sardianos2020rehab}. 

Sensing devices play an essential role in capturing data, and safely storing them in the platform database. To this end, in this article, we focus on investigating various architecture platforms attached to sensors \cite{sardianos2020model}. They are used for uploading wirelessly gathered data from different cubicles to the (EM)\textsuperscript{3} database server that is located at the Qatar university (QU) energy lab. A NoSQL CouchDB server database is deployed to store consumers' micro-moments and occupancy patterns, user preferences and properties, and energy efficiency recommendations and its rating score \cite{alsalemi_rtdpcc_2019,sardianos2020data}. The NoSQL database type was chosen for its fast data retrieval and its flexibility in data structure when compared with traditional SQL-based databases.

The recommendation engine is based on an algorithm that considers user preferences, energy goals, and availability in order to maximize the acceptance of a recommended action and increase the efficiency of the recommender system \cite{himeur2020appliance}. The algorithm is based on the extracted user's habits that concern the repeated usage of devices at certain moments during the day \cite{Alsalemi2020sca}. It is extracted from the energy consumption data and the room occupancy information recorded in users’ (or office) recent history of activities \cite{AymanGPECOM2020}.

Fig. \ref{fig:em3} portrays the overall architecture of (EM)\textsuperscript{3} energy efficiency ecosystem. It is worth noting that the power consumption of the selected devices is considered small.

\begin{figure}[!ht]
\centering
\includegraphics[trim={1.7in 1.7in 1.7in 1.7in},clip,width=\linewidth]{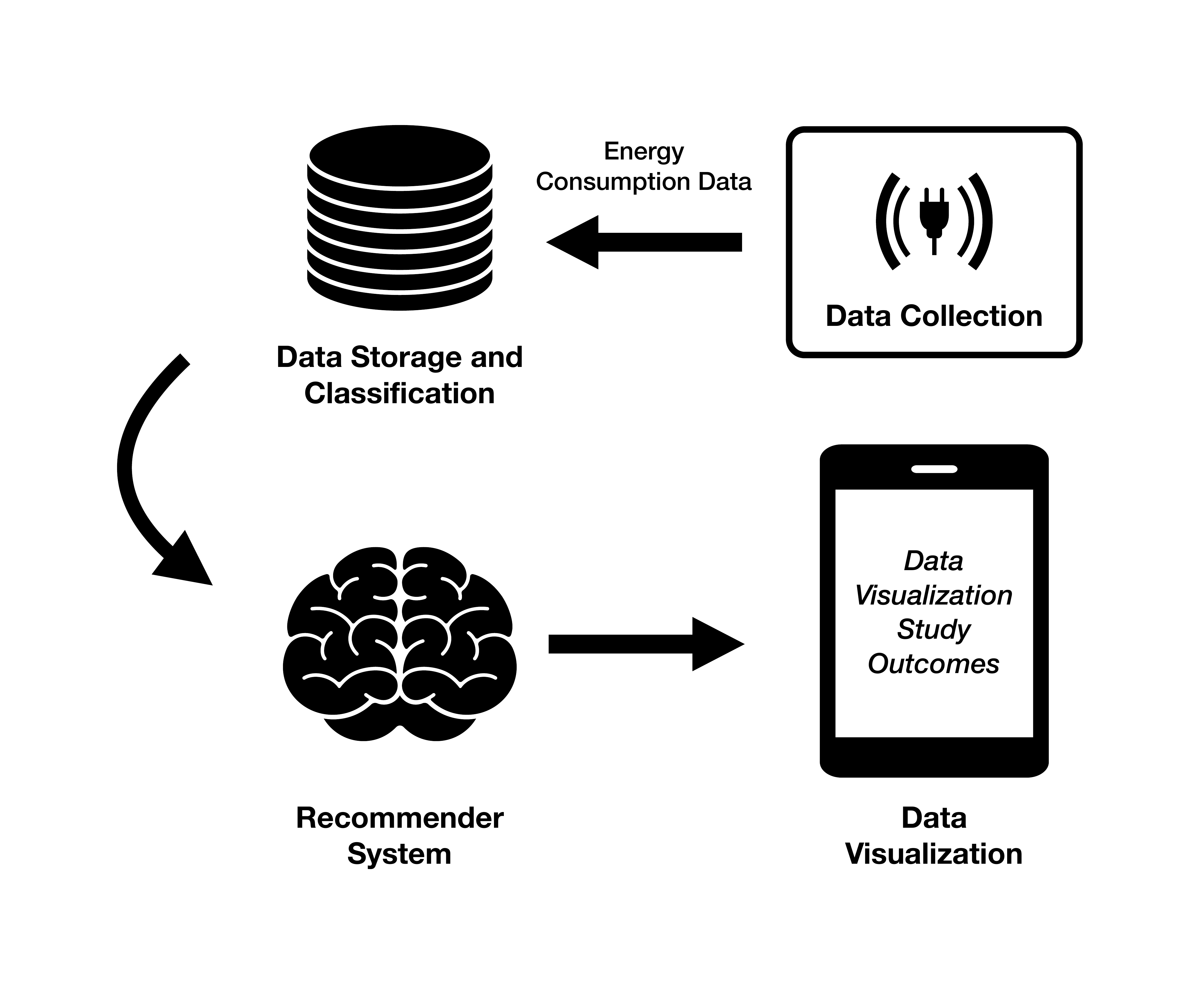}
\caption{Overview of the (EM)\textsuperscript{3} framework.}
\label{fig:em3} 
\end{figure}

The next section describes the selected cloud platforms used for the micro-moment classification phase.

\section{Cloud Evaluation Platforms} \label{sec:platforms}

In order to choose the most suitable platform for cloud classification, a number of criteria is set. First, the platform has to include an accessible interface that is familiar with data scientists, i.e. compatible with common ML programming languages, such as Python and R. Second, the platform shall have different computational power configurations to benchmark the best performance for the algorithm and dataset at hand. Third, from an economical point of view, the platform has to allow researchers to use its functionalities for free to some extent. Based on these criteria, we have selected the following four cloud artificial intelligence platforms:

\begin{itemize}
\item Amazon Web Services Sagemaker (AWSS)\footnote{https://aws.amazon.com/sagemaker}
\item Google Colab (GCL)\footnote{https://Colab.research.google.com}
\item Google Cloud Platform (GCP)\footnote{https://cloud.google.com/products\#ai-and-machine-learning}
\item Microsoft Azure Machine Learning (MAML)\footnote{https://azure.microsoft.com/en-us/services/machine-learning}
\end{itemize}

The above platforms, AWSS, GCL, GCP, and MAML, share a common feature set, which includes Python (or Jupyter notebooks) support, a free plan with limited computational resources, the ability to visualize some of the outcomes of the code run, and the privilege of selecting from numerous computational configurations. Some of the platforms, namely GCP, accept exported TensorFlow models for algorithm execution. 

It goes without saying how big Google services have become and the amount of services that they provide. One of these services is Compute Engine. GCP and GCL offer this service to allow customers to create Virtual Machines (VMs) via “Instances” on Google infrastructure to compute any amount of data. They promise the ability to run thousands of virtual Central Processing Units (vCPUs) quickly with a consistent performance \cite{noauthor_compute_nodate}. Moreover, they provide different machine types with various amounts of vCPUs and memory per vCPU to serve certain purposes \cite{noauthor_machine_nodate}. Not only that, but they also show the specifications of utilized vCPU, and in which machine types they exist \cite{noauthor_cpu_nodate}. Lastly, different NVIDIA GPUs are also highlighted, where they can be added to the created Instances along with the utilized vCPUs \cite{noauthor_gpus_nodate}. Naturally, the variety of options and the amount of heavy computational power they provide do not come for free. In other words, the more computational power (number of vCPUs, GPUs, and memory) is harnessed, the more it will cost the customer. Luckily, Google created Instances in a fashion where the customer can start and stop created Instances, hence, computational payments can only stack-up when the Instances are running. In addition to that, Google provides auto-scalability, where they utilize more Instances only when the traffic is high, and lay off some Instances when the traffic is low. This feature can be harnessed when the customer creates a Managed Instance Group (another feature) for a certain application, where once the traffic is high, more Instances get utilized \cite{noauthor_using_nodate}.

Similar to GCP, MAML allows for Instances to be created to harness the VMs provided by them, similarly to AWSS. Moreover, autoscaling is also a feature that is available to increase the number of Instances when the demand is high, and reduce it when it is low to save customers from paying extra money \cite{rboucher_autoscale_nodate}. This requires the structuring of extra rules for the service to know when to incorporate extra Instances. It is worth mentioning that although the platforms were tested using Jupyter notebooks, they also provide support for Python through an existing Software Development Kit (SDK) for MAML and APIs, and libraries for GCP. In fact, MAML set October 9th, 2020 the day they will retire Azure Notebooks and support plugins to be used with Jupyter notebooks \cite{noauthor_microsoft_nodate}.

It is worth noting that, from the user-experience point-of-view, it was slightly easier to get the first model to run on GCP with respect to its peer MAML. Moreover, for free tier users, it is easier to create and delete Instances on GCP. Although both allow for free trial phase for a whole year, GCP grants users 300 USD to be used in this year, while MAML grants free-tier users 200 USD to be used within the first 30 days of this trial, which is also a plus point for GCP when the customer is an individual, a small business or even a start-up company. Both cloud platforms require a billing card to be registered to ensure that the customer is an authentic user, and to avoid abuses from any potential customers \cite{noauthor_pricing_nodate}. This can similarly apply to CGL. The discussed aspects of the chosen platforms are summarized in Table \ref{tab:scalability}.

\begin{table}[!ht]
\centering
\caption{Cloud platforms scalability comparison}
\label{tab:scalability}
\begin{tabular}{|>{\centering\arraybackslash}m{1.2cm}|>{\centering\arraybackslash}m{1.4cm}|>{\centering\arraybackslash}m{1.3cm}|>{\centering\arraybackslash}m{1.35cm}|>{\centering\arraybackslash}m{1.45cm}|} 
\hline
Platform & Supports Big Data & CPU \,Addition & GPU\, Addition & Scalability \\ 
\hline
GCP & Yes & Yes & Yes & Yes \\ 
\hline
MAML & Yes & Yes & Yes & Yes \\ 
\hline
AWSS & Yes & Yes & Yes & Yes \\
\hline
GCL & Yes & Yes & Yes & No \\ 
\hline
\end{tabular}
\end{table}

In the next section, a description of the utilized datasets for micro-moment classification is provided, which are evaluated within the selected cloud platforms.

\section{Datasets Overview} \label{sec:datasets}

In order to execute a number of classification algorithms to identify micro-moments, relevant datasets are required. They must include appliance-level data points in a household environment. In this work, we have selected the following datasets for cloud classification purposes:

\begin{itemize}
\item \textbf{SimDataset}: The virtual energy dataset (SimDataset), generated by our computer simulator, produces appliance-related datasets based on real data recordings \cite{ramadan_simulator_2019, ALSALEMI2019classifier}. By combining real smart meter data and periodic energy consumption patterns, we simulated sensible domestic electricity consumption scenarios with the aid of k-means clustering, a-priori extraction algorithm, and the innovative use of micro-moments.
\item \textbf{DRED}: The Dutch Residential Energy Dataset (DRED) collected electricity use measurements \cite{uttama2015loced}, occupation trends and ambient evidence of one household in the Netherlands. Sensor systems have been installed to calculate aggregated energy usage and power consumption of appliances. In addition, 12 separate domestic appliances were sub-metered at sampling intervals of 1 min, while 1 Hz sampling rate was used to capture aggregated consumption.
\item \textbf{QUD}: A specific anomaly detection dataset with its ground-truth labels is created on the basis of an experimental setup undertaken at the QU Lab, and is named Qatar University Dataset (QUD) \cite{alsalemi_access_2020, alsalemi2020micro}. A real-time micro-moment facility has been setup to gather reliable data on energy use. The QUD is a collection of readings from different mounted devices (e.g. light lamp, air conditioning, refrigerator, and computer) coupled with quantitative details, such as temperature, humidity, ambient light intensity, and space occupation \cite{himeur2020anomaly}. To the best of the researchers' understanding, QUD is the first dataset in the Middle East in which a normal 240V voltage is used with variable recording duration ranging from 3 seconds to 3 hours \cite{himeur2020building}. 
\end{itemize}

With the aforementioned datasets, varying from simulated, small-scale, and large-scale, cloud artificial intelligence platforms will be utilized to classify those datasets into the following micro-moment classes \cite{ALSALEMI2019classifier}:

\begin{itemize}
\item 0:	good consumption
\item 1:	switch the appliance on
\item 2:	switch the appliance off
\item 3:	excessive power consumption
\item 4:	consumption of power while outside room
\end{itemize}

Next we discuss the equipped ML algorithms to further enhance the understanding of the obtained results. 

\section{Implemented Algorithms} \label{sec:algorthims}
In this work, with selected datasets and cloud platforms, a set of common yet powerful classification algorithms are employed, namely Support Vector Machines (SVM), K-Nearest Neighbors (KNN), and Deep Neural Network (DNN). 

The classification model of SVM is used based on the principle of systemic risk minimisation. This seeks to obtain an optimal isolation hyperplane, which reduces the distance between the features of the same set of appliances. Unless the function trends cannot be segregated linearly in the initial space, the data element can be converted into a new space with higher dimensions by utilizing kernel modules.

In addition, the KNN algorithm is used to distinguish device function characteristics, this algorithm measures the distance of a candidate vector element to identify the $K$ nearest neighbors. The labels are analyzed and used to influence the class label of the candidate feature vector based on the majority vote, and thus, assign a class label to the respective appliance.

Additionally, a novel DNN algorithm is used to classify phenomena. Typically speaking, deep learning is a sub-discipline of ML focused on the concept of studying various degrees of representation by the creation of a hierarchy of characteristics extracted by stacked layers. Keeping this in mind, the DNN system is based on the extension of conventional neural networks by adding additional hidden layers into the network layout between the input and output layers. This is achieved in order to provide a strong capacity to work with dynamic and non-linear grouping issues. As a consequence, DNN has attracted the interest of scientists over the last few years on the ground that it can provide better efficiency than many other current approaches in particular for regression, grouping, simulation, and forecasting goals.

Under this framework, since non-linear separable data are being handled, deep learning is highly recommended for this problem. Furthermore, the efficiency of a deep learning algorithm is typically improved by growing the volume of data used for preparation.

The above algorithms are easily exploited on the selected cloud platforms as Python supports various ML algorithms and these platforms employ Python-based scripts. The yielded results, using the selected datasets, are reported and discussed next. The algorithms are implemented using Python with help of both SciKit Learn and TensorFlow.

\section{Results and Discussion}
\label{sec:results}

This section elaborates on the results of the cloud classification benchmark study. We highlight the performance of each evaluated cloud platform with respect to both the used algorithm and the utilized dataset. Following, light is shed on the limitations and future prospects of cloud artificial intelligence.

\begin{table*}[ht]
\centering
\caption{Cloud classification performance}
\label{tab:results}
\begin{tabular}
{|m{4cm}
|>{\centering\arraybackslash}m{1.8cm}
|>{\centering\arraybackslash}m{1.8cm}
|>{\centering\arraybackslash}m{1.8cm}
|>{\centering\arraybackslash}m{1.8cm}
|>{\centering\arraybackslash}m{1.8cm}
|>{\centering\arraybackslash}m{1.8cm}|}
\hline\hline
\multicolumn{7}{c}{\textbf{DRED}} \\ 
\hline\hline
ML Algorithm & \multicolumn{2}{c|}{SVM} &
\multicolumn{2}{c|}{KNN} &
\multicolumn{2}{c|}{DNN}\\ 
\hline
\backslashbox{Platform}{Performance} & Training Time (s) & Testing Time (s) & Training Time (s) & Testing Time (s) & Training Time (s) & Testing Time (s) \\ 
\hline
MAML & 149.2108 & 12.7986 & 30.6957 & 3.3313 & 515.0723 & 0.2458 \\ 
\hline
GCP & 134.7939 & 10.3981 & 30.0155 & 3.2382 & 1003.7682 & 0.6592 \\ 
\hline
GCL & 84.5881 & 9.2814 & 28.4508 & 3.4481 & 418.3415 & 0.5036 \\ 
\hline
AWSS & 98.8008 & 8.7878 & 31.4675 & 3.2993 & 1102.1146 & 0.7518 \\ 
\hline\hline
\multicolumn{7}{c}{\textbf{QUD}} \\ 
\hline\hline
ML Algorithm & \multicolumn{2}{c|}{SVM} &
\multicolumn{2}{c|}{KNN} &
\multicolumn{2}{c|}{DNN} \\ 
\hline
\backslashbox{Platform}{Performance} & Training Time (s) & Testing Time (s) & Training Time (s) & Testing Time (s) & Training Time (s) & Testing Time (s) \\ 
\hline
MAML & 43.2996 & 4.1559 & 2.1384 & 0.4313 & 142.6328 & 0.0736 \\ 
\hline
GCP & 38.5514 & 3.7396 & 2.2601 & 0.4481 & 291.2316 & 0.2148 \\ 
\hline
GCL & 36.2452 & 3.1886 & 2.7495 & 0.4975 & 122.2153 & 0.1658 \\ 
\hline
AWSS & 34.0591 & 2.9403 & 2.1803 & 0.4050 & 313.4656 & 0.2229 \\ 
\hline\hline
\multicolumn{7}{c}{\textbf{SimDataset}} \\ 
\hline\hline
ML Algorithm & \multicolumn{2}{c|}{SVM} &
\multicolumn{2}{c|}{KNN} &
\multicolumn{2}{c|}{DNN}\\ 
\hline
\backslashbox{Platform}{Performance} & Training Time (s) & Testing Time (s) & Training Time (s) & Testing Time (s) & Training Time (s) & Testing Time (s) \\ 
\hline
MAML & 606.0671 & 28.3848 & 18.6040 & 2.0032 & 318.0821 & 0.1608 \\ 
\hline
GCP & 535.6282 & 25.7717 & 17.7830 & 1.8951 & 643.8005 & 0.4350 \\ 
\hline
GCL & 591.2706 & 23.4997 & 15.5310 & 1.9957 & 281.6197 & 0.3288 \\ 
\hline
AWSS & 572.9653 & 21.4601 & 18.7907 & 2.0166 & 705.4971 & 0.4254 \\
\hline
\end{tabular}
\end{table*}

Table \ref{tab:results} summarizes the classification performance according to the used platform, employed algorithm, and utilized dataset. It is evident that the ML algorithms exhibit varying performance. However, classification on the cloud provides higher performance without burdening the used local hardware. The results are an average computed from three different computation trials. Also, for each algorithm, accuracy and F-score values were similar and were excluded to focus on performance. The used cloud configurations are depicted in Table \ref{tab:conf}.

\begin{table}
\centering
\caption{Used cloud platform configurations}
\label{tab:conf}
\begin{tabular}{|l|l|} 
\hline
Platform & Configuration~          \\ 
\hline
Azure~      & Azure-Standard-D12-v2-28GB~  \\ 
\hline
GCP~        & GCP-n1-highcpu-4-3.60GB       \\ 
\hline
GCL~        & GCL-2-core Xeon-2.2GHz-13GB \\ 
\hline
AWSS~       & AWSS-ml.t3.medium-2vCPU-2GB \\
\hline
\end{tabular}
\end{table}

It is worthy to mention that the used platform exhibited similar performance comparatively considering the free plan option. Both MAML and GCP provided excellent performance, especially for testing. On the other hand, DNN consumed considerably longer for training. This can be explained by the nature of the the neural network, which is highly accelerated and well compensated at testing and deployment phases.

Overall, cloud classification presents an ambitious prospect for ML, especially when local hardware cannot do the job. Embedded systems and Internet of Things (IoT) devices can be considered big users of such platforms. Also, when highly intensive computations are needed, cloud platforms are considered a convenient and economical solution.

\section{Conclusions}
\label{sec:conclusion}
In this paper, common cloud artificial intelligence platforms are benchmarked and compared for micro-moment energy data classification. The AWSS, GCP, GCL, and MAML platforms are tested on the DRED, SimDataset, and QUD datasets. The KNN, DNN, and SVM classifiers have been employed. Superb performance has been observed in the cloud platform showing relatively close performance. Yet, the nature of some algorithms limits the training performance, such as DNN. Future work includes evaluating more platforms and integrating with the energy efficiency (EM)\textsuperscript{3} framework.

\section*{Acknowledgment}
\label{acknowledgements}
This paper was made possible by National Priorities Research Program (NPRP) grant No. 10-0130-170288 from the Qatar National Research Fund (a member of Qatar Foundation). The statements made herein are solely the responsibility of the authors.

%\bibliographystyle{IEEEtran}
%\bibliography{RTDPCC2020}

\begin{thebibliography}{10}

\bibitem{dupont_defusing_2020}
C.~Dupont, ``Defusing contested authority: {EU} energy efficiency
  policymaking,'' \emph{Journal of European Integration}, vol.~42, no.~1, pp.
  95--110, Jan. 2020. [Online]. Available:
  \url{https://doi.org/10.1080/07036337.2019.1708346}


\bibitem{langsdorf2011eu}
S.~Langsdorf, ``Eu energy policy: from the ecsc to the energy roadmap 2050,''
  \emph{Brussels: Green European Foundation}, 2011.

\bibitem{himeur2020efficient}
Y.~Himeur, A.~Alsalemi, F.~Bensaali, and A.~Amira, ``Efficient multi-descriptor
  fusion for non-intrusive appliance recognition,'' in \emph{2020 IEEE
  International Symposium on Circuits and Systems (ISCAS)}.\hskip 1em plus
  0.5em minus 0.4em\relax IEEE, 2020, pp. 1--5.

\bibitem{Himeur2020IJIS-NILM}
Y.~{Himeur}, A.~{Elsalemi}, F.~{Bensaali}, and A.~Amira, ``An intelligent
  non-intrusive load monitoring scheme based on 2d phase encoding of power
  signals,'' \emph{Intenational Journal of Intelligent Systems}, pp. 1--22,
  2020.

\bibitem{HIMEUR2020115872}
Y.~Himeur, A.~Alsalemi, F.~Bensaali, and A.~Amira, ``Effective non-intrusive
  load monitoring of buildings based on a novel multi-descriptor fusion with
  dimensionality reduction,'' \emph{Applied Energy}, vol. 279, p. 115872, 2020.

\bibitem{de2008smart}
J.~De~Las~Heras and A.~Zarli, ``The smart buildings group report on ict for
  energy efficiency,'' \emph{Final Report, ICT for Energy Efficiency Ad-hoc
  Advisory Group}, 2008.

\bibitem{Varlamis2020CCIS}
I.~Varlamis, C.~Sardianos, G.~Dimitrakopoulos, A.~Alsalemi, Y.~Himeur,
  F.~Bensaali, and A.~Amira, ``Reshaping consumption habits by exploiting
  energy-related micro-moment recommendations: A case study,'' in
  \emph{Communications in Computer and Information Science}.\hskip 1em plus
  0.5em minus 0.4em\relax Cham: Springer International Publishing, 2020, pp.
  1--22.

\bibitem{moranreducing}
A.~J. Mor{\'a}n, P.~Profaizer, M.~H. Zapater, and I.~Z. Bribi{\'a}n, ``Reducing
  energy consumption in buildings with information and communication
  technologies (icts)--technology review and analysis of results from eu pilot
  projects.''

\bibitem{hannus2010ict}
M.~Hannus, A.~S. Kazi \emph{et~al.}, ``Ict supported energy efficiency in
  construction: strategic research roadmap and implementation
  recommendations,'' 2010.

\bibitem{ye2008ict}
J.~Ye, T.~Hassan, C.~Carter, and A.~Zarli, ``Ict for energy efficiency: The
  case for smart buildings,'' \emph{Department of Civil and Building
  Engineering, Loughborough University}, 2008.

\bibitem{horner2016known}
N.~C. Horner, A.~Shehabi, and I.~L. Azevedo, ``Known unknowns: indirect energy
  effects of information and communication technology,'' \emph{Environmental
  Research Letters}, vol.~11, no.~10, p. 103001, 2016.

\bibitem{himeur2020improving}
Y.~{Himeur}, A.~{Elsalemi}, F.~{Bensaali}, and A.~Amira, ``Improving in-home
  appliance identification using fuzzy-neighbors-preserving analysis based
  qr-decomposition,'' pp. 1--8, Fabruary 2020.

\bibitem{Sardianos2020IJIS-ERS}
C.~Sardianos, I.~Varlamis, G.~Dimitrakopoulos, D.~Anagnostopoulo, A.~Alsalemi,
  Y.~Himeur, F.~Bensaali, and A.~Amira, ``The emergence of explainability of
  intelligent systems: Delivering explainable and personalised recommendations
  for energy efficiency,'' \emph{Intenational Journal of Intelligent Systems},
  pp. 1--22, 2020.

\bibitem{Sardianos2020GreenCom}
C.~{Sardianos}, C.~{Chronis}, I.~{Varlamis}, G.~{Dimitrakopoulos}, Y.~{Himeur},
  A.~{Alsalemi}, F.~{Bensaali}, and A.~{Amira}, ``Real-time personalised energy
  saving recommendations,'' in \emph{The 16th IEEE International Conference on
  Green Computing and Communications (GreenCom)}, 2020, pp. 1--6.

\bibitem{alsalemi2019ieeesystems}
A.~{Alsalemi}, C.~{Sardianos}, F.~{Bensaali}, I.~{Varlamis}, A.~{Amira}, and
  G.~{Dimitrakopoulos}, ``The role of micro-moments: A survey of habitual
  behavior change and recommender systems for energy saving,'' \emph{IEEE
  Systems Journal}, pp. 1--12, 2019.

\bibitem{ALSALEMI2019classifier}
A.~Alsalemi, M.~Ramadan, F.~Bensaali, A.~Amira, C.~Sardianos, I.~Varlamis, and
  G.~Dimitrakopoulos, ``Endorsing domestic energy saving behavior using
  micro-moment classification,'' \emph{Applied Energy}, vol. 250, pp. 1302 --
  1311, 2019.

\bibitem{patel2016wearable}
S.~Patel, R.~S. McGinnis, I.~Silva, S.~DiCristofaro, N.~Mahadevan, E.~Jortberg,
  J.~Franco, A.~Martin, J.~Lust, M.~Raj \emph{et~al.}, ``A wearable computing
  platform for developing cloud-based machine learning models for health
  monitoring applications,'' in \emph{2016 38th Annual International Conference
  of the IEEE Engineering in Medicine and Biology Society (EMBC)}.\hskip 1em
  plus 0.5em minus 0.4em\relax IEEE, 2016, pp. 5997--6001.

\bibitem{bihis2015generalized}
M.~Bihis and S.~Roychowdhury, ``A generalized flow for multi-class and binary
  classification tasks: An azure ml approach,'' in \emph{2015 IEEE
  International Conference on Big Data (Big Data)}.\hskip 1em plus 0.5em minus
  0.4em\relax IEEE, 2015, pp. 1728--1737.

\bibitem{roychowdhury2016ag}
S.~Roychowdhury and M.~Bihis, ``Ag-mic: Azure-based generalized flow for
  medical image classification,'' \emph{IEEE Access}, vol.~4, pp. 5243--5257,
  2016.

\bibitem{alsalemi_access_2020}
A.~{Alsalemi}, Y.~{Himeur}, F.~{Bensaali}, A.~{Amira}, C.~{Sardianos},
  I.~{Varlamis}, and G.~{Dimitrakopoulos}, ``Achieving domestic energy
  efficiency using micro-moments and intelligent recommendations,'' \emph{IEEE
  Access}, vol.~8, pp. 15\,047--15\,055, 2020.

\bibitem{alsalemi_rtdpcc_2019}
A.~Alsalemi, M.~Ramadan, F.~Bensaali, A.~Amira, C.~Sardianos, I.~Varlamis, and
  G.~Dimitrakopoulos, ``{Boosting {Domestic} {Energy}
  {Efficiency} {Through} {Accurate} {Consumption} {Data} {Collection}},''
  Leicester, UK, 2019.

\bibitem{alsalemi2020micro}
A.~Alsalemi, Y.~Himeur, F.~Bensaali, A.~Amira, C.~Sardianos, C.~Chronis,
  I.~Varlamis, and G.~Dimitrakopoulos, ``A micro-moment system for domestic
  energy efficiency analysis,'' \emph{IEEE Systems Journal}, 2020.

\bibitem{himeur2020novel}
Y.~Himeur, A.~Alsalemi, F.~Bensaali, and A.~Amira, ``A novel approach for
  detecting anomalous energy consumption based on micro-moments and deep neural
  networks,'' \emph{Cognitive Computation}, pp. 1--21, 2020.

\bibitem{himeur2020applicability}
Y.~Himeur, A.~Alsalemi, F.~Bensaali, A.~Amira, C.~Sardianos, I.~Varlamis, and
  G.~Dimitrakopoulos, ``On the applicability of 2d local binary patterns for
  identifying electrical appliances in non-intrusive load monitoring,'' in
  \emph{Proceedings of SAI Intelligent Systems Conference}.\hskip 1em plus
  0.5em minus 0.4em\relax Springer, 2020, pp. 188--205.

\bibitem{himeur2020data}
Y.~Himeur, A.~Alsalemi, A.~Al-Kababji, F.~Bensaali, and A.~Amira, ``Data fusion
  strategies for energy efficiency in buildings: Overview, challenges and novel
  orientations,'' \emph{Information Fusion}, vol.~64, pp. 99--120, 2020.

\bibitem{himeur2020robust}
Y.~Himeur, A.~Alsalemi, F.~Bensaali, and A.~Amira, ``Robust event-based
  non-intrusive appliance recognition using multi-scale wavelet packet tree and
  ensemble bagging tree,'' \emph{Applied Energy}, vol. 267, p. 114877, 2020.

\bibitem{sardianos_smartgreens_2019}
C.~Sardianos, I.~Varlamis, G.~Dimitrakopoulos, D.~Anagnostopoulos, A.~Alsalemi,
  F.~Bensaali, and A.~Amira, ``"{{I} {Want} to ...
  {Change}": {Micro}-moment based {Recommendations} can {Change} {Users}'
  {Energy} {Habits}},'' Heraklion, Crete - Greece, 2019, pp. 30--39.

\bibitem{sardianos2020rehab}
C.~Sardianos, I.~Varlamis, G.~Dimitrakopoulos, D.~Anagnostopoulos, A.~Alsalemi,
  F.~Bensaali, Y.~Himeur, and A.~Amira, ``Rehab-c: Recommendations for energy
  habits change,'' \emph{Future Generation Computer Systems}, vol. 112, pp. 394
  -- 407, 2020.

\bibitem{sardianos2020model}
C.~Sardianos, I.~Varlamis, C.~Chronis, G.~Dimitrakopoulos, Y.~Himeur,
  A.~Alsalemi, F.~Bensaali, and A.~Amira, ``A model for predicting room
  occupancy based on motion sensor data,'' in \emph{2020 IEEE International
  Conference on Informatics, IoT, and Enabling Technologies (ICIoT)}.\hskip 1em
  plus 0.5em minus 0.4em\relax IEEE, 2020, pp. 394--399.

\bibitem{sardianos2020data}
C.~{Sardianos}, I.~{Varlamis}, C.~{Chronis}, G.~{Dimitrakopoulos}, Y.~{Himeur},
  A.~{Alsalemi}, F.~{Bensaali}, and A.~{Amira}, ``Data analytics, automations,
  and micro-moment based recommendations for energy efficiency,'' in \emph{2020
  IEEE Sixth International Conference on Big Data Computing Service and
  Applications (BigDataService)}, 2020, pp. 96--103.

\bibitem{himeur2020appliance}
Y.~{Himeur}, A.~{Elsalemi}, F.~{Bensaali}, and A.~Amira, ``Appliance
  identification using a histogram post-processing of 2d local binary patterns
  for smart grid applications,'' pp. 1--8, May 2020.

\bibitem{Alsalemi2020sca}
A.~{Elsalemi}, Y.~{Himeur}, F.~{Bensaali}, and A.~Amira, ``Appliance-level
  monitoring with micro-moment smart plugs,'' in \emph{The Fifth International
  Conference on Smart City Applications (SCA)}, May 2020, pp. 1--5.

\bibitem{AymanGPECOM2020}
A.~Al-Kababji, A.~Alsalemi, Y.~Himeur, R.~fernandez, F.~Bensaali, A.~Amira, and
  N.~Fetais, ``Energy data visualizations on smartphones for triggering
  behavioral change: Novel vs. conventional,'' in \emph{The 2nd of Global
  Power, Energy and Communication Conference (GPECOM)}, 2020, pp. 1--6.

\bibitem{noauthor_compute_nodate}
``{Compute {Engine} documentation {\textbar} {Compute}
  {Engine} {Documentation}}.'' [Online]. Available:
  \url{https://cloud.google.com/compute/docs}

\bibitem{noauthor_machine_nodate}
``{Machine types {\textbar} {Compute} {Engine}
  {Documentation}}.'' [Online]. Available:
  \url{https://cloud.google.com/compute/docs/machine-types}

\bibitem{noauthor_cpu_nodate}
``{{CPU} platforms {\textbar} {Compute} {Engine}
  {Documentation}}.'' [Online]. Available:
  \url{https://cloud.google.com/compute/docs/cpu-platforms}

\bibitem{noauthor_gpus_nodate}
``{{GPUs} on {Compute} {Engine} {\textbar} {Compute}
  {Engine} {Documentation}}.'' [Online]. Available:
  \url{https://cloud.google.com/compute/docs/gpus}


\bibitem{noauthor_using_nodate}
``{Using autoscaling for highly scalable
  applications}.'' [Online]. Available:
  \url{https://cloud.google.com/compute/docs/tutorials/high-scalability-autoscaling}

\bibitem{rboucher_autoscale_nodate}
{rboucher}, ``{Autoscale in {Microsoft} {Azure} -
  {Azure} {Monitor}}.'' [Online]. Available:
  \url{https://docs.microsoft.com/en-us/azure/azure-monitor/platform/autoscale-overview}

\bibitem{noauthor_microsoft_nodate}
``Microsoft {Azure} {Notebooks}.'' [Online]. Available:
  \url{https://notebooks.azure.com/Content/alternatives.html}


\bibitem{noauthor_pricing_nodate}
``{Pricing {Calculator} {\textbar} {Microsoft}
  {Azure}}.'' [Online]. Available:
  \url{https://azure.microsoft.com/en-us/pricing/calculator/}

\bibitem{ramadan_simulator_2019}
M.~Ramadan, A.~Alsalemi, F.~Bensaali, A.~Amira, C.~Sardianos, I.~Varlamis,
  G.~Dimitrakopoulos, and D.~Anagnostopoulos,
  ``{Simulating {Appliance}-{Based} {Power}
  {Consumption} {Records} for {Energy} {Efficiency} {Awareness}},'' Västerås,
  Sweden, 2019.

\bibitem{uttama2015loced}
A.~S. Uttama~Nambi, A.~Reyes~Lua, and V.~R. Prasad, ``Loced: Location-aware
  energy disaggregation framework,'' in \emph{Proceedings of the 2nd acm
  international conference on embedded systems for energy-efficient built
  environments}, 2015, pp. 45--54.

\bibitem{himeur2020anomaly}
Y.~Himeur, K.~Ghanem, A.~Alsalemi, F.~Bensaali, and A.~Amira, ``Anomaly
  detection of energy consumption in buildings: A review, current trends and
  new perspectives,'' \emph{arXiv preprint arXiv:2010.04560}, 2020.

\bibitem{himeur2020building}
Y.~{Himeur}, A.~{Elsalemi}, F.~{Bensaali}, and A.~Amira, ``Building power
  consumption datasets: Survey, taxonomy and future directions,'' \emph{Energy
  and Buildings}, vol. 227, p. 110404, 2020.

\end{thebibliography}
%


\end{document}